\documentclass[prl,twocolumn,superscriptaddress,floatfix,longbibliography]{revtex4-2}

\usepackage{graphicx}
\usepackage{amsmath,amssymb,bm}
\usepackage{hyperref}

\newcommand{\ii}{\mathrm{i}} 

\graphicspath{{Figures/}{./}}

\begin{document}

\title{Constant-Amplitude $2\pi$ Phase Modulation from Topological Pole--Zero Winding}

\author{Alex Krasnok}
\email{akrasnok@fiu.edu}
\affiliation{Department of Electrical Engineering, Florida International University, Miami, Florida 33174, USA}
\affiliation{Knight Foundation School of Computing and Information Sciences, Florida International University, Miami, Florida 33174, USA}

\begin{abstract}
Resonant phase shifters inevitably mix phase and amplitude. We present a topological synthesis that
guarantees a full $2\pi$ phase swing at a prescribed constant scattering magnitude $|S_{ij}|=C$ by winding
a scattering zero around the operating point in the complex-frequency plane while avoiding pole windings.
We realize this either by complex-frequency waveform excitation on an iso-$|S_{ij}|$ (Apollonius) loop or by
adiabatic co-modulation of detuning and decay at fixed carrier, suppressing AM--PM conversion and
quantizing $\Delta\phi$ by the Argument Principle. The approach targets integrated resonant modulators, programmable photonic circuits, and quantum/beam-steering
interferometers that require amplitude-flat phase shifts.
\end{abstract}

\maketitle

\textit{Introduction.—}
Optical phase control underpins coherent modulation formats in telecommunications, interferometric metrology, beam steering, and reconfigurable photonic processors \cite{ref1,ref2,ref3,ref4,ref5,ref6,ref7}.
In integrated platforms, phase control is commonly achieved via refractive-index tuning (thermo-optic, electro-optic, carrier dispersion, phase-change materials), often in resonant geometries to reduce footprint and drive power \cite{ref1,ref4,ref14,ref15,ref16,ref18}.
A persistent limitation is \emph{parasitic amplitude modulation} that accompanies resonant phase tuning and degrades high-fidelity operations (AM--PM conversion) \cite{ref12,ref13,ref17}.
Moreover, realizing a full $2\pi$ shift can require large parameter excursions or increased dissipation, especially under tight integration constraints \cite{ref14,ref16,ref18}.
While single-bus microrings are widely used as all-pass/notch elements with steep phase response, material loss and imperfect coupling move practical devices away from unitary scattering and reintroduce amplitude--phase entanglement \cite{Bogaerts2012, Sacher2013}.
These constraints motivate protocols that decouple phase from amplitude while retaining a full $2\pi$ range within compact resonant platforms.

\begin{figure}[!b]
\centering
\includegraphics[width=0.35\textwidth]{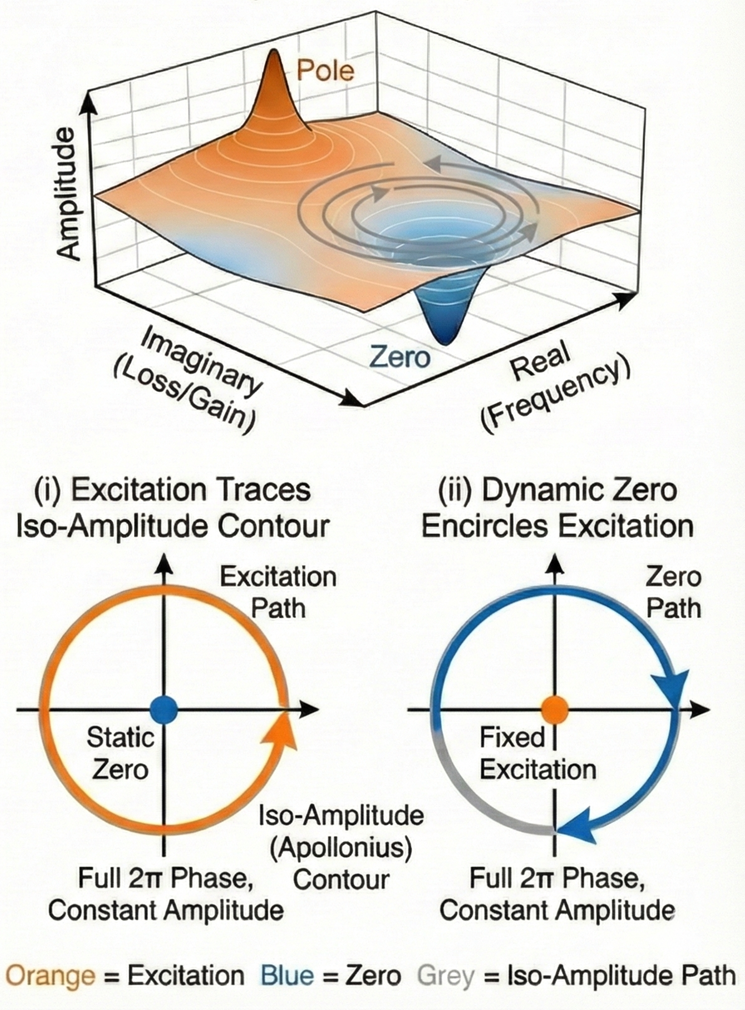}
\caption{Concept of topological phase control in the complex-frequency plane.
A resonant response landscape is characterized by a pole (peak) and a zero (dip).
A closed trajectory constrained to an iso-$|S_{ij}|$ manifold and enclosing a net topological charge produces a quantized $2\pi$ phase accumulation while enforcing constant amplitude.
Protocol (i) uses a waveform-synthesized complex-frequency excitation to trace an iso-amplitude loop around a static zero; Protocol (ii) co-modulates device parameters to move the zero around a fixed real drive while maintaining $|S_{ij}|=C$.}
\label{fig:concept}
\end{figure}

Recent progress in non-Hermitian and topological photonics emphasizes that linear responses are organized by their poles and zeros in the complex-frequency plane \cite{ref19,ref20,ref21,ref22,ref23,ref25}.
Phase winding around these critical points provides quantized control primitives and has been leveraged in scattering singularities and metasurface phase engineering \cite{ref24,ref59Kim2025,ref27,ref28}.
More generally, for linear time-invariant photonic networks the relevant scattering coefficients (e.g., reflection/transmission matrix elements) are meromorphic functions of complex frequency, a structure made explicit within multiport temporal coupled-mode theory \cite{ref50,ref51}.
In this Letter, we exploit a simple, general observation: if the operating point traces a closed contour that encloses a response zero but not a pole, the net phase accumulation is fixed to $2\pi$ by topology (Cauchy’s Argument Principle) \cite{ref29,ref30,ref31}.
By additionally constraining motion to an \emph{iso-amplitude} contour, one obtains pure phase modulation at constant amplitude.
The novelty of this Letter is to elevate this topological statement into a \emph{prescriptive synthesis rule} for realistic (lossy) resonators and arbitrary scattering channels: for a target level $C$, we explicitly construct a closed iso-$|S_{ij}|$ loop that encloses a net topological charge $(N_0-N_p)$, thereby guaranteeing $\Delta\phi=2\pi(N_0-N_p)$ while maintaining $|S_{ij}|=C$ \emph{by construction}.
This is distinct from conventional all-pass conditions (which require near-unitarity) and from interferometric amplitude equalization: the constant-amplitude constraint is enforced at the level of the selected scattering coefficient itself, and the $2\pi$ phase swing is topologically quantized and therefore robust to continuous loop deformations that preserve the enclosed critical points.
Approaching a scattering zero enables a given phase action with a smaller tuning excursion, but the reduced signal magnitude can penalize SNR and increase phase-noise sensitivity, motivating multiport routing or interferometric recombination when needed. We connect both protocols (Fig.~\ref{fig:concept}, below) to established experimental knobs: waveform-shaped drives that emulate complex-frequency excitation (virtual critical coupling) \cite{ref57Younes2020,ref58Hinney2024,ref59Kim2025} and independent control of coupling and dissipation in integrated resonators \cite{Sacher2013,ref53,ref54}, including coupling-modulated microring platforms \cite{Sacher2013}. Throughout, we assume adiabatic modulation, $\Omega_{\rm mod}\ll 2\pi(\Gamma_0+\Gamma_c)$, so the scattering tracks the instantaneous pole--zero constellation and Floquet sidebands are negligible; faster modulation requires a full time-varying/Floquet scattering treatment \cite{ref61}.

\textit{Model and complex-plane structure.—}
Any \emph{linear, time-invariant, causal} photonic structure that is interfaced through well-defined input/output channels is characterized by a frequency-domain scattering matrix $\bm S(\tilde f)$ relating incoming to outgoing complex wave amplitudes. Causality implies that $\bm S(\tilde f)$ is analytic for $\mathrm{Im}\{\tilde f\}>0$ under our time convention $e^{\ii 2\pi \tilde f t}$, while passivity and reciprocity further constrain its analytic continuation in the complex-frequency plane. In open resonant systems, the singularities of $\bm S$ coincide with the complex eigenfrequencies of the underlying leaky modes (quasinormal modes), enabling reduced-order descriptions in terms of a small set of poles and zeros \cite{Alpeggiani2017}.
Within temporal coupled-mode theory (TCMT), when a \emph{single isolated resonance} dominates a given scattering channel, each matrix element can be cast (up to a slowly varying background $S_{ij}^{\rm bg}$) into a first-order \emph{M\"obius} form with one pole $\tilde f_p$ and a channel-dependent zero $\tilde f_{z,ij}$ \cite{ref50,ref51}:
\begin{equation}
S_{ij}(\tilde f)=S_{ij}^{\rm bg}\,\frac{\tilde f-\tilde f_{z,ij}}{\tilde f-\tilde f_p}.
\label{eq:mobius}
\end{equation}
Equation~(\ref{eq:mobius}) makes explicit the central geometric point of this Letter: \emph{constant-amplitude phase control} can be synthesized by prescribing a closed contour in the $\tilde f$-plane that (i) remains on an iso-$|S_{ij}|$ manifold and (ii) encloses a net topological charge $(N_0-N_p)$, which quantizes the accumulated phase.

\begin{figure}[!t]
\centering
\includegraphics[width=0.4\textwidth]{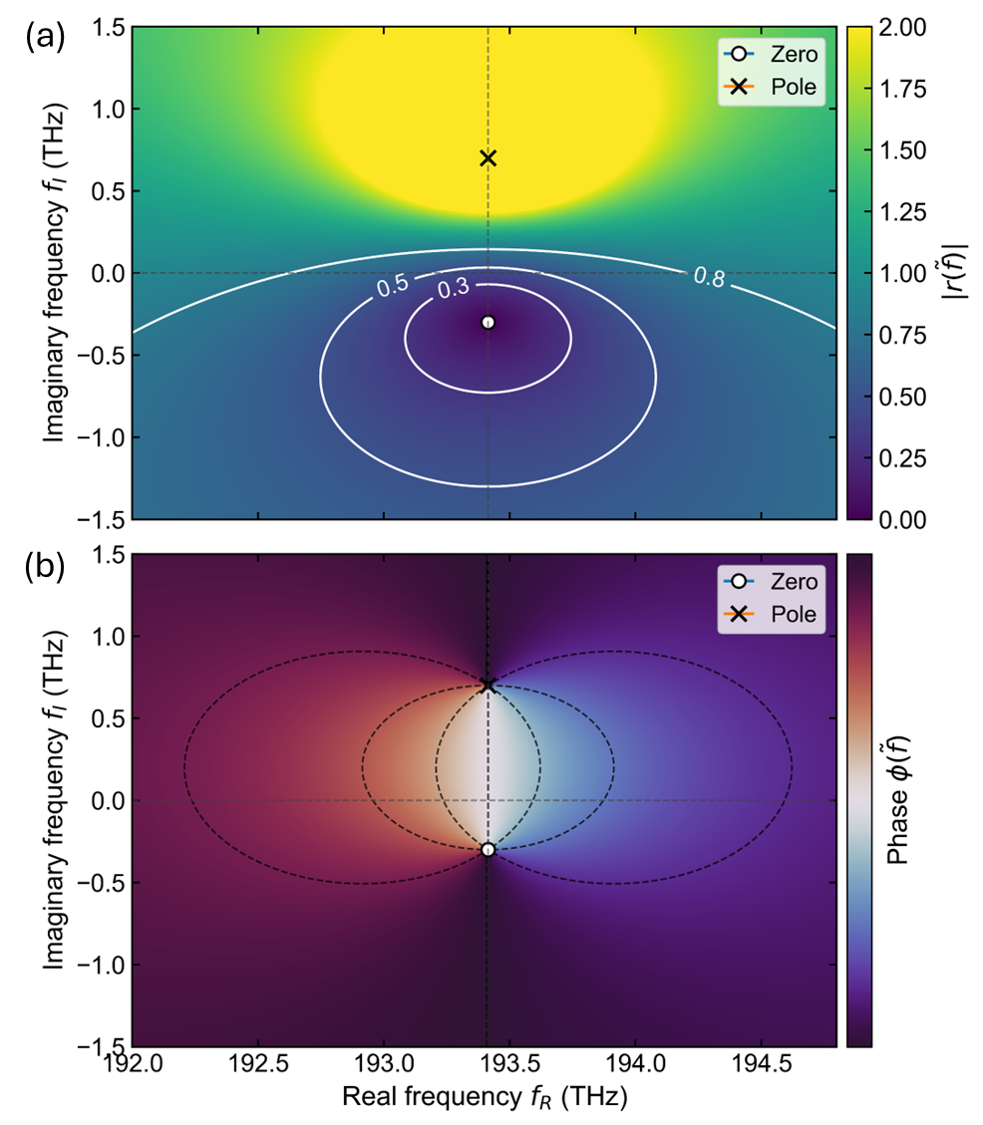}
\caption{Complex-frequency maps of the reflection coefficient $r(\tilde f)$ for an illustrative parameter set $f_0\simeq 193.41~\mathrm{THz}$ ($\lambda_0\simeq 1.55~\mu\mathrm{m}$), $\Gamma_0=0.2~\mathrm{THz}$, and $\Gamma_c=0.5~\mathrm{THz}$ (overcoupled). (a) Amplitude $|r(\tilde f)|$ with the zero $\tilde f_z=f_0+\ii(\Gamma_0-\Gamma_c)\simeq 193.41-\ii 0.3~\mathrm{THz}$ and pole $\tilde f_p=f_0+\ii(\Gamma_0+\Gamma_c)\simeq 193.41+\ii 0.7~\mathrm{THz}$ marked; white curves show representative iso-$|r|$ (Apollonius) contours used for constant-amplitude phase winding. (b) Phase $\phi(\tilde f)$ exhibiting $+2\pi$ winding around the zero and $-2\pi$ winding around the pole.}
\label{fig:maps}
\end{figure}

To isolate the pole--zero geometry without extraneous multiport bookkeeping, we specialize to the canonical case of a single resonator side-coupled to a waveguide and treat the reflected field as an effective one-port response. Extension to multiport devices follows directly by replacing $r$ with the desired scattering element $S_{ij}$ in TCMT \cite{ref50,ref51}.
In the complex-frequency plane $\tilde f=f_R+\ii f_I$, the reflection coefficient can be written as
\begin{equation}
r(\tilde f)=\frac{\ii(\tilde f-f_0)+(\Gamma_0-\Gamma_c)}{\ii(\tilde f-f_0)+(\Gamma_0+\Gamma_c)} ,
\label{eq:r}
\end{equation}
where $f_0$ is the resonance frequency, $\Gamma_0$ the intrinsic loss rate, and $\Gamma_c$ the external coupling rate.
Equivalently, Eq.~(\ref{eq:r}) factorizes into an explicit pole--zero ratio $r(\tilde f)=(\tilde f-\tilde f_z)/(\tilde f-\tilde f_p)$, with a reflection zero and pole at $\tilde f_z = f_0+\ii(\Gamma_0-\Gamma_c)$, $\tilde f_p = f_0+\ii(\Gamma_0+\Gamma_c)$. This structure connects directly to widely used physical interpretations of scattering zeros and poles in open systems (e.g., moving a scattering zero onto the real axis under appropriate dissipation and coherent driving) \cite{Chong2010}.

The phase of $r$ satisfies the Argument Principle: for a closed contour $C$ in the $\tilde f$-plane,
\begin{equation}
\Delta\phi_C = 2\pi\,(N_0-N_p),
\label{eq:argprin}
\end{equation}
where $N_0$ ($N_p$) is the number of enclosed zeros (poles) of $r$ \cite{ref29,ref30,ref31}.
Hence, any closed trajectory that encloses $\tilde f_z$ but excludes $\tilde f_p$ yields a topologically quantized phase accumulation $\Delta\phi_C=2\pi$, independent of smooth deformations of the path that do not cross singularities.

Crucially, we seek loops that maintain constant amplitude $|r(\tilde f)|=C$.
Using the factorized form, the constant-amplitude constraint becomes a purely geometric condition $|\tilde f-\tilde f_z|=C\,|\tilde f-\tilde f_p|$, i.e., the locus of points with fixed distance ratio to two foci $(\tilde f_z,\tilde f_p)$---an \emph{Apollonius circle} (a standard consequence of M\"obius geometry) \cite{Needham1997}.
For Eq.~(\ref{eq:r}), iso-amplitude contours are therefore Apollonius circles (see \textit{SM Sec.~3} for details),
\begin{equation}
(f_R-f_0)^2+\!\left[f_I-\!\left(\Gamma_0-\Gamma_c\frac{1+C^2}{1-C^2}\right)\right]^2
=\left(\frac{2C\Gamma_c}{|1-C^2|}\right)^2 .
\label{eq:apollonius}
\end{equation}
Equation~(\ref{eq:apollonius}) is the closed-form \emph{synthesis rule} for the constant-amplitude trajectory at a prescribed level $C$: selecting the Apollonius circle that encloses $\tilde f_z$ but not $\tilde f_p$ guarantees \emph{simultaneously} (i) $|r|=C$ and (ii) $\Delta\phi=2\pi$, thereby eliminating AM--PM calibration and rendering the full-range phase action topologically robust.
Figure~\ref{fig:maps} visualizes the amplitude/phase landscapes and representative iso-$|r|$ contours; values at $f_I\neq 0$ correspond to the analytic continuation of the linear response in the complex-frequency (Laplace) domain and should not be interpreted as steady-state gain.

\textit{Approach 1: complex-frequency excitation on an iso-amplitude loop.—}
In the first protocol the device is \emph{time-invariant}: the resonator parameters $(f_0,\Gamma_0,\Gamma_c)$ are fixed, and phase control is achieved by \emph{waveform synthesis} of the incident field so that the excitation follows a designed trajectory $\tilde f_{\rm exc}(t)=f_R(t)+\ii f_I(t)$ in the complex-frequency plane. A constant complex-frequency tone $e^{\ii 2\pi \tilde f t}$ is equivalent to a carrier at $f_R$ with an exponential envelope $e^{-2\pi f_I t}$ (decay for $f_I>0$, growth for $f_I<0$), while a piecewise-smooth $\tilde f_{\rm exc}(t)$ corresponds to joint control of instantaneous frequency and complex-envelope amplitude. Such control is available in optical arbitrary waveform generation and line-by-line pulse shaping platforms \cite{CundiffWeiner2010,Weiner2011,Huang2008,Jiang2005}.
Segments with $f_I<0$ imply transient growth and therefore require a finite temporal window and bounded dynamic range; this is precisely the experimentally validated setting of complex-frequency loading (virtual critical coupling), where tailored pulses emulate complex-frequency drives to control resonator energy transfer in integrated photonics \cite{ref57Younes2020,ref58Hinney2024}.

Selecting a target amplitude $C$ fixes an iso-$|r|$ contour via Eq.~(\ref{eq:apollonius}). Driving $\tilde f_{\rm exc}(t)$ around the corresponding Apollonius loop that encloses the reflection zero $\tilde f_z$ but not the pole $\tilde f_p$ yields a topologically quantized phase accumulation $\Delta\phi=2\pi$ (Argument Principle), while enforcing $|r(\tilde f_{\rm exc})|=C$ \emph{by construction}. Here ``constant amplitude'' refers to the magnitude of the \emph{scattering coefficient} evaluated on the designed complex-frequency trajectory; the reflected waveform inherits the programmed input envelope. If a constant output-envelope is required at the system level, one may normalize by the known input envelope or embed the element in a multiport/interferometric architecture without changing the winding-based phase guarantee.

\begin{figure}[t]
\centering
\includegraphics[width=0.4\textwidth]{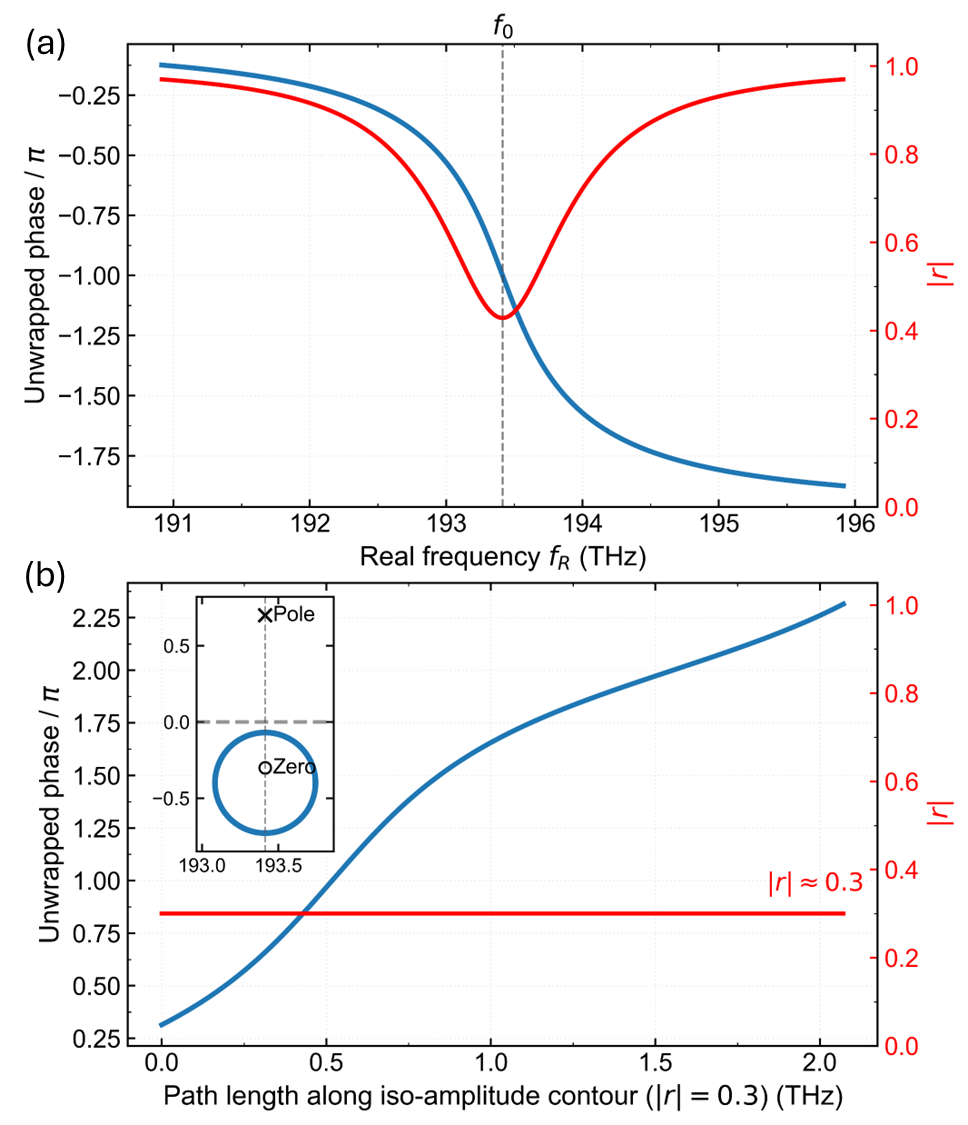}
\caption{Approach 1: phase winding at constant amplitude by steering the excitation along a complex-frequency iso-$|r|$ contour. (a) Real-axis sweep ($f_I=0$) exhibits an incomplete phase excursion and significant amplitude variation. (b) A synthesized complex-frequency drive $\tilde f_{\mathrm{exc}}(t)$ traverses an Apollonius contour ($|r|=0.3$) enclosing $\tilde f_z$ (excluding $\tilde f_p$, inset), yielding a $2\pi$ unwrapped phase shift while maintaining constant $|r|$ (within numerical tolerance).}
\label{fig:app1}
\end{figure}

Figure~\ref{fig:app1} contrasts a conventional real-frequency sweep [$f_I=0$, Fig.~\ref{fig:app1}(a)], which produces an incomplete phase excursion accompanied by strong amplitude variation (AM--PM coupling), with an iso-amplitude loop at $|r|=0.3$, which realizes a full $2\pi$ unwrapped phase shift while maintaining $|r|$ fixed, Fig.~\ref{fig:app1}(b). The same construction applies to any scattering channel by replacing $r$ with the desired $S_{ij}$. For more details on complex-frequency excitation (CFE) as finite-time waveform synthesis and dynamic range, see \textit{SM Sec. 5}.

\textit{Approach 2: dynamic pole--zero motion around a fixed-frequency excitation.—}
The second protocol keeps the drive \emph{strictly monochromatic and real}: a continuous-wave excitation at $f_{\mathrm{exc}}$ is applied, and phase-only control is obtained by \emph{adiabatically} steering the instantaneous pole--zero constellation relative to this fixed operating point.
In the adiabatic regime, $\Omega_{\mathrm{mod}}\ll 2\pi(\Gamma_0+\Gamma_c)$, the carrier response follows the instantaneous scattering coefficient; at higher speeds, Floquet sidebands and non-adiabatic distortions enter and require a full time-varying scattering treatment \cite{Galiffi2022}.
At a fixed $f_{\mathrm{exc}}$, the time-dependent reflection can be expressed in pole--zero form as
$r(f_{\mathrm{exc}},t)=\big(f_{\mathrm{exc}}-\tilde f_z(t)\big)/\big(f_{\mathrm{exc}}-\tilde f_p(t)\big)$,
with $\tilde f_z(t)=f_0(t)+\ii(\Gamma_0(t)-\Gamma_c)$ and $\tilde f_p(t)=f_0(t)+\ii(\Gamma_0(t)+\Gamma_c)$.
If $\tilde f_z(t)$ executes one counterclockwise winding around the point $f_{\mathrm{exc}}$ while $\tilde f_p(t)$ does not, the Argument Principle guarantees a quantized phase accumulation $\Delta\phi=2\pi$ at the operating frequency.
To suppress AM--PM conversion simultaneously, we enforce the \emph{iso-amplitude} constraint $|r(f_{\mathrm{exc}},t)|=C_d$ over the entire modulation cycle.

\begin{figure}[!t]
\centering
\includegraphics[width=0.5\textwidth]{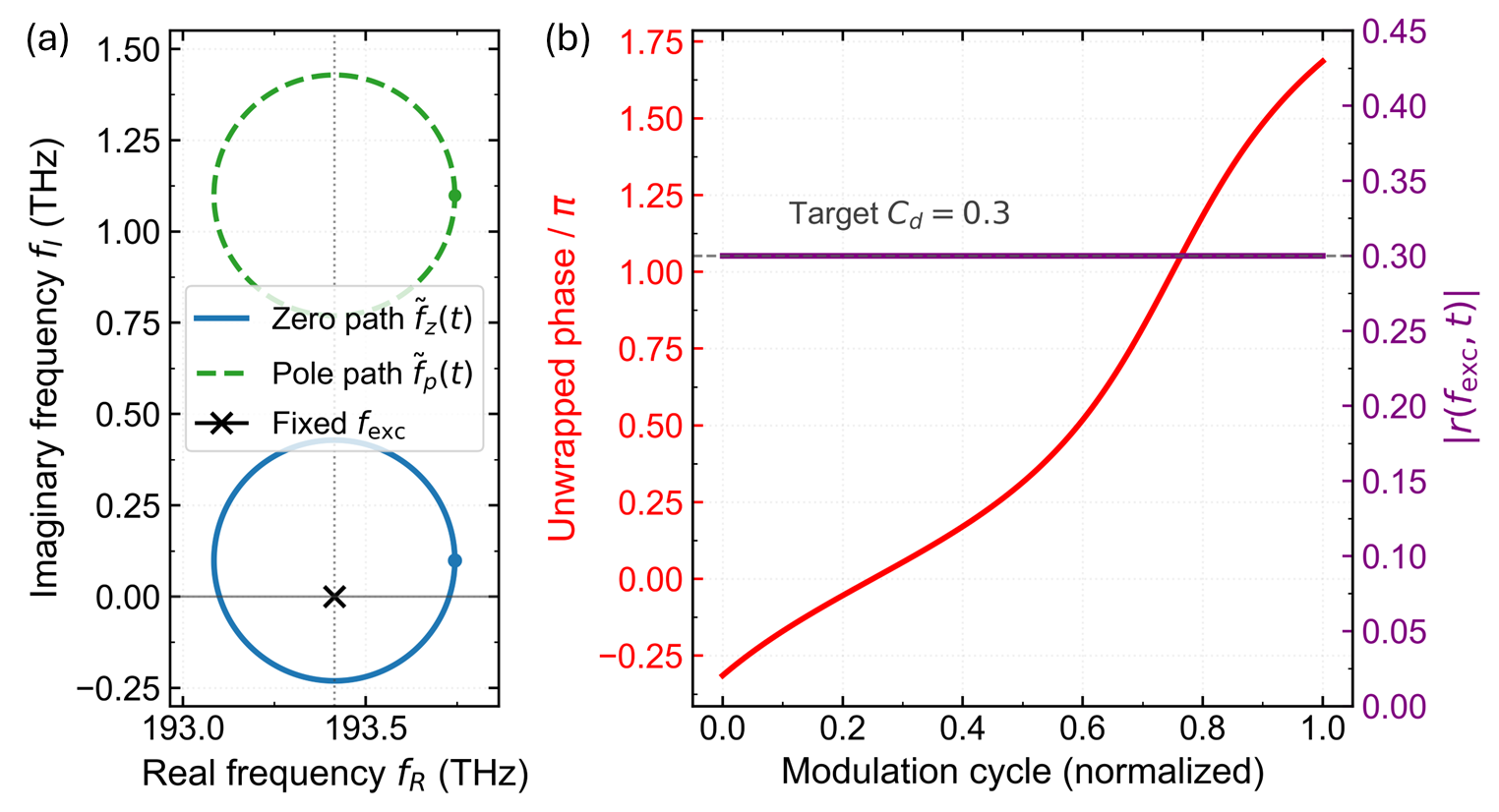}
\caption{Approach 2: phase winding at constant amplitude with a fixed real excitation. (a) Trajectories of the dynamically tuned zero $\tilde f_z(t)$ (solid) and pole $\tilde f_p(t)$ (dashed) in the complex plane; the fixed excitation $f_{\mathrm{exc}}$ is enclosed by the zero’s path but not the pole’s. (b) At $f_{\mathrm{exc}}$, the unwrapped phase accumulates $2\pi$ over one modulation cycle while $|r(f_{\mathrm{exc}},t)|$ is held at the target $C_d$ (within numerical tolerance).}
\label{fig:app2}
\end{figure}

We modulate $f_0(t)$ and $\Gamma_0(t)$ (with $\Gamma_c$ fixed) to enforce
$|r(f_{\mathrm{exc}},t)|=C_d$.
Geometrically, this requires that $(f_0(t),\Gamma_0(t))$ trace a closed loop that keeps $f_{\mathrm{exc}}$ on a constant-$|r|$ manifold of the \emph{instantaneous} system, while producing a nontrivial winding of $\tilde f_z(t)$ about $f_{\mathrm{exc}}$.
The iso-amplitude constraint yields a circular trajectory in the $(f_0,\Gamma_0)$ parameter space, with $\Gamma_{0,\mathrm{center}} = \frac{1+C_d^2}{1-C_d^2}\,\Gamma_c$, $R_{\mathrm{param}} = \frac{2C_d}{1-C_d^2}\,\Gamma_c.$ 
For $C_d=0.3$ and $\Gamma_c=0.5~\mathrm{THz}$, $R_{\mathrm{param}}\simeq 0.33~\mathrm{THz}$, see \textit{SM Sec.~4} for details.
Because $\tilde f_p(t)$ is vertically offset from $\tilde f_z(t)$ by $2\Gamma_c$, this construction naturally yields loops where the zero encloses $f_{\mathrm{exc}}$ while the pole remains non-enclosing (Fig.~\ref{fig:app2}), producing a robust $2\pi$ winding at fixed amplitude (within numerical tolerance).
In practice, $f_0(t)$ can be tuned by electro-optic or thermo-optic actuation, while the effective damping can be tuned via controlled absorption (e.g., carrier-induced or graphene-assisted loss) or via a tunable coupling to an auxiliary dissipative channel \cite{ref53,ref58}.
Alternatively, one may co-modulate detuning and \emph{external} coupling $\Gamma_c(t)$, which has been demonstrated in coupling-modulated microrings (including operation beyond the linewidth limit), relaxing the need for direct intrinsic-loss control \cite{Sacher2013}.

As the target amplitude is reduced ($C_d\!\downarrow$), the required zero motion and the parameter-space excursion shrink (SM Sec.~S8), so weaker tuning mechanisms can, in principle, deliver a full $2\pi$ phase action.
The unavoidable trade-off is reduced raw signal and increased phase-noise sensitivity as $|r|\!\to\!0$; in practice this can be mitigated by interferometric recombination or multiport routing while preserving the topologically quantized winding set by the enclosed singularities.

\textit{Discussion and outlook.—}
The two protocols above provide a compact \emph{synthesis rule} for phase-only control that is both constructive and topologically protected: choose a target level $|S_{ij}|=C$, then implement a closed trajectory on the corresponding iso-amplitude manifold that encloses the relevant scattering zero(s) but not the pole(s), guaranteeing a quantized phase accumulation $\Delta\phi=2\pi(N_0-N_p)$ while suppressing AM--PM conversion by construction.
This robustness is qualitatively different from ``near-constant-amplitude'' phase shifters that rely on fine balancing of loss and coupling: here the \emph{net} phase winding is fixed by the enclosed singularities, and the amplitude is fixed by the chosen iso-$|S_{ij}|$ contour, enabling an \emph{arbitrary} user-defined constant magnitude even in lossy, non-unitary devices.

Approach~1 shifts complexity to the source: waveform-synthesized drives can realize the required complex-frequency trajectories using optical arbitrary waveform generation and line-by-line pulse shaping, and complex-frequency loading has been experimentally validated in the context of virtual critical coupling \cite{CundiffWeiner2010,ref57Younes2020}.
Because segments with $f_I<0$ imply transient growth, implementations are naturally finite-duration and dynamic-range limited; nevertheless, the \emph{winding-based} phase guarantee remains intact so long as the realized path preserves the enclosed critical-point content.
Approach~2 instead uses a standard continuous-wave drive at fixed $f_{\rm exc}$ and moves the burden to device-side tuning, where detuning, coupling, and effective damping can be controlled with mature integrated-actuation toolkits; in particular, fast external-coupling control has been demonstrated in coupling-modulated microrings, including operation beyond the linewidth limit \cite{Sacher2013}.

Our analysis assumes adiabatic modulation, $\Omega_{\rm mod}\ll 2\pi(\Gamma_0+\Gamma_c)$, under which the carrier response follows the instantaneous pole--zero constellation. Extending the synthesis to non-adiabatic regimes (with sideband generation and Floquet scattering) is a promising direction for future work \cite{Galiffi2022}. On adiabaticity and side-band generation for time-modulated parameters, see \textit{SM. Sec. 6.} Finally, while operating closer to a scattering zero reduces the required tuning excursion for a given phase action, it also reduces raw signal and increases phase-noise susceptibility as $|S_{ij}|\!\to\!0$; practical systems can mitigate this trade-off via interferometric recombination or multiport routing while preserving the topologically quantized winding set by the enclosed singularities.
Because the construction relies only on the analytic structure of scattering coefficients, it applies directly to transmission-phase control in two-port filters and to selected channels of larger multiport networks by replacing $r$ with the desired $S_{ij}$.

\textit{Conclusions.—}
We introduced a topological, pole--zero based synthesis for complete $2\pi$ phase control at a prescribed constant scattering amplitude. By enforcing an iso-$|S_{ij}|$ constraint and a nontrivial zero winding without pole winding, we obtain phase-only modulation with suppressed AM--PM conversion and a quantized net phase set by the Argument Principle.
Two complementary realizations—complex-frequency waveform excitation and adiabatic device-side tuning at fixed carrier—provide practical pathways compatible with modern photonic control capabilities. The resulting design paradigm is general across scattering channels and multiport networks, and offers a route to drift-tolerant, calibration-light phase control for coherent communications, programmable photonics, and quantum photonic circuits.

\begin{acknowledgments}
The author acknowledges financial support from the U.S. Department of Energy (DoE) and the U.S. Air Force Office of Scientific Research (AFOSR).
\end{acknowledgments}

\bibliographystyle{apsrev4-2}
\bibliography{references}

@article{ref1,
  author  = {Parra, J. and Navarro-Arenas, J. and Sanchis, P.},
  title   = {Silicon thermo-optic phase shifters: a review of configurations and optimization strategies},
  journal = {Advanced Photonics Nexus},
  volume  = {3},
  pages   = {044001},
  year    = {2024}
}

@article{ref2,
  author  = {Xu, T. and Dong, Y. and Zhong, Q. and Zheng, S. and Qiu, Y. and Zhao, X. and Jia, L. and Lee, C. and Hu, T.},
  title   = {Mid-infrared integrated electro-optic modulators: a review},
  journal = {Nanophotonics},
  volume  = {12},
  pages   = {3683},
  year    = {2023}
}

@book{ref3,
  author    = {Kawanishi, T.},
  title     = {Electro-Optic Modulation for Photonic Networks: Precise and High-Speed Control of Lightwaves},
  publisher = {Springer},
  address   = {Cham},
  year      = {2022}
}

@article{ref4,
  author  = {Rahim, A. and Hermans, A. and Wohlfeil, B. and Petousi, D. and Kuyken, B. and Van Thourhout, D. and Baets, R.},
  title   = {Taking silicon photonics modulators to a higher performance level: state-of-the-art and a review of new technologies},
  journal = {Advanced Photonics},
  volume  = {3},
  pages   = {024003},
  year    = {2021}
}

@article{ref5,
  author  = {Xu, M. and others},
  title   = {High-performance coherent optical modulators based on thin-film lithium niobate platform},
  journal = {Nature Communications},
  volume  = {11},
  pages   = {3911},
  year    = {2020}
}

@article{ref6,
  author  = {Reines, I. C. and Wood, M. G. and Luk, T. S. and Serkland, D. K. and Campione, S.},
  title   = {Compact epsilon-near-zero silicon photonic phase modulators},
  journal = {Optics Express},
  volume  = {26},
  pages   = {21594},
  year    = {2018}
}

@article{ref7,
  author  = {Dong, C. and others},
  title   = {An ultra-compact integrated phase shifter via electrically tunable meta-waveguides},
  journal = {Nanoscale Horizons},
  volume  = {10},
  pages   = {933},
  year    = {2025}
}

@inproceedings{ref12,
  author       = {Taylor, J. A. and Quinlan, F. and Hati, A. and Nelson, C. and Diddams, S. A. and Datta, S. and Joshi, A.},
  title        = {Phase Noise in the Photodetection of Ultrashort Optical Pulses},
  booktitle    = {2010 IEEE International Frequency Control Symposium},
  organization = {IEEE},
  address      = {Newport Beach, CA, USA},
  pages        = {684--688},
  year         = {2010}
}

@article{ref13,
  author  = {Tai, Z. and Yan, L. and Zhang, Y. and Zhang, X. and Guo, W. and Zhang, S. and Jiang, H.},
  title   = {Electro-optic modulator with ultra-low residual amplitude modulation for frequency modulation and laser stabilization},
  journal = {Optics Letters},
  volume  = {41},
  pages   = {5584},
  year    = {2016}
}

@article{ref14,
  author  = {Sun, H. and Qiao, Q. and Guan, Q. and Zhou, G.},
  title   = {Silicon Photonic Phase Shifters and Their Applications: A Review},
  journal = {Micromachines},
  volume  = {13},
  pages   = {1509},
  year    = {2022}
}

@article{ref15,
  author  = {Prabhathan, P. and others},
  title   = {Roadmap for phase change materials in photonics and beyond},
  journal = {iScience},
  volume  = {26},
  pages   = {107946},
  year    = {2023}
}

@article{ref16,
  author  = {Harris, N. C. and Ma, Y. and Mower, J. and Baehr-Jones, T. and Englund, D. and Hochberg, M. and Galland, C.},
  title   = {Efficient, compact and low loss thermo-optic phase shifter in silicon},
  journal = {Optics Express},
  volume  = {22},
  pages   = {10487},
  year    = {2014}
}

@article{ref17,
  author  = {Kincaid, P. S. and Andriolli, N. and Contestabile, G. and De Marinis, L.},
  title   = {Addressing optical modulator non-linearities for photonic neural networks},
  journal = {Communications Engineering},
  volume  = {4},
  pages   = {58},
  year    = {2025}
}

@article{ref18,
  author  = {Kim, J. Y. and Park, J. and Holdman, G. R. and Heiden, J. T. and Kim, S. and Brar, V. W. and Jang, M. S.},
  title   = {Full $2\pi$ tunable phase modulation using avoided crossing of resonances},
  journal = {Nature Communications},
  volume  = {13},
  pages   = {2103},
  year    = {2022}
}

@article{ref19,
  author  = {Ashida, Y. and Gong, Z. and Ueda, M.},
  title   = {Non-Hermitian physics},
  journal = {Advances in Physics},
  volume  = {69},
  pages   = {249},
  year    = {2020}
}

@article{ref20,
  author  = {Bender, C. M. M.},
  title   = {Making sense of non-Hermitian Hamiltonians},
  journal = {Reports on Progress in Physics},
  volume  = {70},
  pages   = {947},
  year    = {2007}
}

@article{ref21,
  author  = {Bergholtz, E. J. and Budich, J. C. and Kunst, F. K.},
  title   = {Exceptional topology of non-Hermitian systems},
  journal = {Reviews of Modern Physics},
  volume  = {93},
  pages   = {015005},
  year    = {2021}
}

@article{ref22,
  author  = {Feng, L. and El-Ganainy, R. and Ge, L.},
  title   = {Non-Hermitian photonics based on parity-time symmetry},
  journal = {Nature Photonics},
  volume  = {11},
  pages   = {752},
  year    = {2017}
}

@article{ref23,
  author  = {Miri, M.-A. and Al\`u, A.},
  title   = {Exceptional points in optics and photonics},
  journal = {Science},
  volume  = {363},
  pages   = {eaar7709},
  year    = {2019}
}

@article{ref24,
  author  = {Sakotic, Z. and Krasnok, A. and Al\`u, A. and Jankovic, N.},
  title   = {Topological scattering singularities and embedded eigenstates for polarization control and sensing applications},
  journal = {Photonics Research},
  volume  = {9},
  pages   = {1310},
  year    = {2021}
}

@article{ref25,
  author  = {Gong, Z. and Ashida, Y. and Kawabata, K. and Takasan, K. and Higashikawa, S. and Ueda, M.},
  title   = {Topological Phases of Non-Hermitian Systems},
  journal = {Physical Review X},
  volume  = {8},
  pages   = {031079},
  year    = {2018}
}

@article{ref27,
  author  = {Krasnok, A. and Baranov, D. and Li, H. and Miri, M.-A. and Monticone, F. and Al\`u, A.},
  title   = {Anomalies in light scattering},
  journal = {Advances in Optics and Photonics},
  volume  = {11},
  pages   = {892},
  year    = {2019}
}

@article{ref28,
  author  = {Colom, R. and Mikheeva, E. and Achouri, K. and Zuniga-Perez, J. and Bonod, N. and Martin, O. J. F. and Burger, S. and Genevet, P.},
  title   = {Crossing of the Branch Cut: The Topological Origin of a Universal $2\pi$-Phase Retardation in Non-Hermitian Metasurfaces},
  journal = {Laser \& Photonics Reviews},
  volume  = {17},
  pages   = {2200976},
  year    = {2023}
}

@book{ref29,
  author    = {Ahlfors, L. V.},
  title     = {Complex Analysis: An Introduction to the Theory of Analytic Functions of One Complex Variable},
  publisher = {American Mathematical Society},
  address   = {Providence, Rhode Island},
  year      = {2021},
  edition   = {3}
}

@inproceedings{ref30,
  author    = {Li, W. and Paulson, L. C.},
  title     = {A Formal Proof of Cauchy’s Residue Theorem},
  booktitle = {Interactive Theorem Proving},
  editor    = {Blanchette, J. C. and Merz, S.},
  volume    = {9807},
  publisher = {Springer International Publishing},
  address   = {Cham},
  pages     = {235--251},
  year      = {2016}
}

@book{ref31,
  author    = {Howie, J. M.},
  title     = {Complex Analysis},
  publisher = {Springer},
  address   = {London},
  year      = {2004},
  note      = {Second printing}
}

@article{ref50,
  author  = {Fan, Shanhui and Suh, Wonjoo and Joannopoulos, J. D.},
  title   = {Temporal coupled-mode theory for the Fano resonance in optical resonators},
  journal = {J. Opt. Soc. Am. A},
  volume  = {20},
  number  = {3},
  pages   = {569--572},
  year    = {2003},
  doi     = {10.1364/JOSAA.20.000569}
}

@article{ref51,
  author  = {Suh, Wonjoo and Wang, Zheng and Fan, Shanhui},
  title   = {Temporal Coupled-Mode Theory and the Presence of Non-Orthogonal Modes in Lossless Multimode Cavities},
  journal = {IEEE J. Quantum Electron.},
  volume  = {40},
  number  = {10},
  pages   = {1511--1518},
  year    = {2004},
  doi     = {10.1109/JQE.2004.834773}
}

@article{ref53,
  author  = {Ding, Yunhong and Zhu, Xiaolong and Xiao, Sanshui and Hu, Hao and Frandsen, Lars Hagedorn and Mortensen, N. Asger and Yvind, Kresten},
  title   = {Effective Electro-Optical Modulation with High Extinction Ratio by a Graphene-Silicon Microring Resonator},
  journal = {Nano Lett.},
  volume  = {15},
  number  = {7},
  pages   = {4393--4400},
  year    = {2015},
  doi     = {10.1021/acs.nanolett.5b00630}
}

@article{ref54,
  author  = {P{\'e}rez-L{\'o}pez, Daniel and Gutierrez, Ana M. and S{\'a}nchez, Erica and DasMahapatra, Prometheus and Capmany, Jos{\'e}},
  title   = {Integrated photonic tunable basic units using dual-drive directional couplers},
  journal = {Opt. Express},
  volume  = {27},
  number  = {26},
  pages   = {38071--38086},
  year    = {2019},
  doi     = {10.1364/OE.27.038071}
}

@article{ref58,
  author  = {Rezzonico, Daniele and Jazbinsek, Mojca and Guarino, Andrea and Kwon, O.-Pil and G{\"u}nter, Peter},
  title   = {Electro-optic Charon polymeric microring modulators},
  journal = {Opt. Express},
  volume  = {16},
  number  = {2},
  pages   = {613--627},
  year    = {2008},
  doi     = {10.1364/OE.16.000613}
}

@article{ref61,
  author  = {Yariv, Amnon},
  title   = {Universal relations for coupling of optical power between microresonators and dielectric waveguides},
  journal = {Electronics Letters},
  volume  = {36},
  number  = {4},
  pages   = {321--322},
  year    = {2000},
  doi     = {10.1049/el:20000340}
}

@article{Sacher2013,
  author  = {Sacher, Wesley D. and Green, William M. J. and Assefa, S. and Barwicz, T. and Pan, H. and Shank, S. M. and Vlasov, Y. A. and Poon, Joyce K. S.},
  title   = {Coupling modulation of microrings at rates beyond the linewidth limit},
  journal = {Optics Express},
  volume  = {21},
  number  = {8},
  pages   = {9722--9733},
  year    = {2013},
  doi     = {10.1364/OE.21.009722}
}

@article{Bogaerts2012,
  author  = {Bogaerts, Wim and De Heyn, Peter and Van Vaerenbergh, Thomas and De Vos, Kris and Selvaraja, Shankar Kumar and Claes, Tom and Dumon, Pieter and Bienstman, Peter and Van Thourhout, Dries and Baets, Roel},
  title   = {Silicon microring resonators},
  journal = {Laser \& Photonics Reviews},
  volume  = {6},
  number  = {1},
  pages   = {47--73},
  year    = {2012},
  doi     = {10.1002/lpor.201100017}
}

@article{Alpeggiani2017,
  author  = {Alpeggiani, Filippo and Parappurath, Nikhil and Verhagen, Ewold and Kuipers, L.},
  title   = {Quasinormal-Mode Expansion of the Scattering Matrix},
  journal = {Physical Review X},
  volume  = {7},
  pages   = {021035},
  year    = {2017},
  doi     = {10.1103/PhysRevX.7.021035}
}

@article{Chong2010,
  author  = {Chong, Y. D. and Ge, Li and Cao, Hui and Stone, A. D.},
  title   = {Coherent Perfect Absorbers: Time-Reversed Lasers},
  journal = {Physical Review Letters},
  volume  = {105},
  pages   = {053901},
  year    = {2010},
  doi     = {10.1103/PhysRevLett.105.053901}
}

@book{Needham1997,
  author    = {Needham, Tristan},
  title     = {Visual Complex Analysis},
  publisher = {Oxford University Press},
  address   = {Oxford},
  year      = {1997}
}

@article{CundiffWeiner2010,
  author  = {Cundiff, Steven T. and Weiner, Andrew M.},
  title   = {Optical arbitrary waveform generation},
  journal = {Nature Photonics},
  volume  = {4},
  pages   = {760--766},
  year    = {2010},
  doi     = {10.1038/nphoton.2010.196}
}

@article{Weiner2011,
  author  = {Weiner, Andrew M.},
  title   = {Ultrafast optical pulse shaping: A tutorial review},
  journal = {Optics Communications},
  volume  = {284},
  pages   = {3669--3692},
  year    = {2011},
  doi     = {10.1016/j.optcom.2011.03.084}
}

@article{Huang2008,
  author  = {Huang, Chen-Bin and Jiang, Zhifang and Leaird, Daniel E. and Caraquitena, Jose and Weiner, Andrew M.},
  title   = {Spectral line-by-line shaping for optical and microwave arbitrary waveform generation},
  journal = {Laser \& Photonics Reviews},
  volume  = {2},
  number  = {4},
  pages   = {227--248},
  year    = {2008},
  doi     = {10.1002/lpor.200810001}
}

@article{Jiang2005,
  author  = {Jiang, Zhifang and Leaird, Daniel E. and Weiner, Andrew M.},
  title   = {Line-by-line pulse shaping control for optical arbitrary waveform generation},
  journal = {Optics Express},
  volume  = {13},
  number  = {25},
  pages   = {10431--10439},
  year    = {2005},
  doi     = {10.1364/OPEX.13.010431}
}

@article{Galiffi2022,
  author  = {Galiffi, Emanuele and Tirole, Romain and Yin, Shixiong and Li, Huanan and Vezzoli, Stefano and Huidobro, Paloma A. and Silveirinha, M{\'a}rio G. and Al{\`u}, Andrea and Pendry, John B.},
  title   = {Photonics of time-varying media},
  journal = {Advanced Photonics},
  volume  = {4},
  number  = {1},
  pages   = {014002},
  year    = {2022},
  doi     = {10.1117/1.AP.4.1.014002}
}

@article{ref57Younes2020,
  author  = {Ra'di, Younes and Krasnok, Alex and Al{\`u}, Andrea},
  title   = {Virtual Critical Coupling},
  journal = {ACS Photonics},
  volume  = {7},
  number  = {6},
  pages   = {1468--1475},
  year    = {2020},
  doi     = {10.1021/acsphotonics.0c00165}
}

@article{ref58Hinney2024,
  author  = {Hinney, Jakob and Kim, Seunghwi and Flatt, Graydon J. K. and Datta, Ipshita and Al{\`u}, Andrea and Lipson, Michal},
  title   = {Efficient excitation and control of integrated photonic circuits with virtual critical coupling},
  journal = {Nature Communications},
  volume  = {15},
  pages   = {2741},
  year    = {2024},
  doi     = {10.1038/s41467-024-46908-2}
}

@article{ref59Kim2025,
  author  = {Kim, Seunghwi and Krasnok, Alex and Al{\`u}, Andrea},
  title   = {Complex-frequency excitations in photonics and wave physics},
  journal = {Science},
  volume  = {387},
  number  = {6741},
  pages   = {eado4128},
  year    = {2025},
  doi     = {10.1126/science.ado4128}
}

\end{document}